\documentclass[a4paper,10pt,3p,preprint,sort&compress,times]{elsarticle}
\biboptions{square,comma}

\usepackage{amsmath}
\usepackage{setspace}
\usepackage{graphicx}
\usepackage{caption}
\usepackage{subfigure}

\begin{document}

\begin{frontmatter}

\title{Efficient routing on small complex networks without buffers}

\author[]{J. Smiljani\'{c}}
\ead{ jelena.smiljanic@ipb.ac.rs }
\author[]{I. Stankovi\'{c}}

\address[]{Scientific Computing Laboratory, Institute of Physics Belgrade, University of Belgrade, Pregrevica 118, 11080 Belgrade, Serbia}

\date{\today, revised version}

\begin{abstract}
In this paper, we are exploring strategies for the reduction of
the congestion in the complex networks. The nodes without buffers
are considered, so, if the congestion occurs, the information
packets will be dropped. The focus is on the efficient routing.
The routing strategies are compared using two generic models,
i.e., Barab\`asi-Albert scale-free network and scale-free network
on lattice, and the academic router networks of the Netherlands
and France. We propose a dynamic deflection routing algorithm
which automatically extends path of the packet before it arrives
at congested node. The simulation results indicate that the
dynamic routing strategy can further reduce number of dropped
packets in a combination with the efficient path routing proposed
by Yan et al. [Phys. Rev. E 73, 046108 (2006)].
\end{abstract}


\end{frontmatter}

\section{Introduction}

Complex networks are important for functioning of the modern
society. To ensure a free, uncongested traffic flows on the
complex networks is of great interest. Intuitively, the traffic
congestion could be largely reduced or completely avoided with a
very large average degree of connectivity and/or node capacity for
information packet delivery. The capacity of nodes to deliver
information cannot be infinite. Also, upgrading the infrastructure
is often not economically feasible~\cite{kim1,kim2}. The
performance of the communication systems can be improved by
implementing the more appropriate routing protocols without
changing the underlying network
structure~\cite{tang,yan,danila1,danila2,
guimera,wang1,yin,sreenivasan,pu,wang2,kujawski,echenique1,echenique2,ling,tintor},
which is more realizable in the practice. Such work presents two
problems. The first is finding out the optimal strategies for the
traffic routing on a defined network structure. The second problem
is finding a procedure to draw general conclusions about
performance of routing strategies due to the variation of the real
network topologies. The increasing speed of the network interfaces
raises an important question concerning the size of buffers,
complexity and cost. A considerable research effort is currently
under way in an attempt to resolve compromise between buffer
latency and complexity on one side and capacity on the
other~\cite{wischik1,wischik2,raina,
enachescu1,enachescu2,gorinsky1,gorinsky2,wong}. In the previous
studies, the node buffer size in the traffic-flow model is set as
infinite~\cite{tang,yan,danila1,danila2,
guimera,wang1,yin,sreenivasan,pu,wang2,kujawski,echenique1,echenique2,ling}.
Our intention is to produce a relatively simple methodology for
evaluating routing strategies in networks with limited buffering
capability or without optical buffers. This should be important
for the optical networks with either large volumes of the
information traffic or without buffering capacity.

A number of network models is introduced in the last two
decade~\cite{BA, dorogovtsev}. A particular class of models is
dedicated to networks embedded in the space~\cite{rozenfeld}. Here
we are interested in evaluating the models for representation of
the spatially constrained networks, both in terms distance between
nodes and the extent of the network. We are interested in
information flow optimization in the small networks. The small
networks should represent bulk of telecommunication networks or
other dedicated information networks, e.g., regional optical
backbones, academic networks. As examples of the the real-world
networks, we analyze the national research and educational
networks (NRENs) of the Netherlands~\cite{surfnet},
France~\cite{renater}, Norway~\cite{uninett} and
Spain~\cite{rediris} and compare them with Barab\`asi-Albert
scale-free network and scale-free network on lattice. The system
size dependence of topological characteristics of the scale-free
network on lattice is also analyzed.

In this work three routing strategies are implemented and
evaluated: the shortest path routing, efficient path routing and
dynamic deflection routing. The shortest path routing is widely
used routing strategy in praxis (by "shortest" we mean the path
with the smallest number of links)~\cite{dijkstra}. However, in
the shortest path routing strategy load distribution is not
homogeneous. The majority of the shortest paths pass through the
nodes that are highly connected, while other nodes carry much less
traffic~\cite{goh}. Yan et al.~\cite{yan} presented an approach to
redistribute traffic load in highly connected nodes to other nodes
using link weight. An improvement is achieved through a targeted
traffic redistribution from the most congested nodes. As result
the congestion is reduced at expense of a slight increase of the
total path length and traffic. We compare this routing strategy
with a dynamic routing strategy. The dynamic routing strategy
improves the control of the congestion in the heavily loaded nodes
by dynamically returning packet one step back. In this way, the
dynamic strategy uses the redundant capacity of the links in
network to temporarily store information, until congested node
capacity is free. Further, we test a possibility of combining
dynamic and static routing strategy. In order to check how
different routing strategies behave in the larger networks, we
have evaluated information loss dependence on packet generation
rate and network size in case of the scale-free network on
lattices.

The paper is organized as follows: in Sec.II we introduce the
generic scale-free model and the scale-free model on lattice and
compare their network characteristics with the national research
and educational networks (NRENs) of the Netherlands, France,
Norway and Spain. Furthermore, the characteristics of the
scale-free model on lattice for different system sizes are
considered. In Sec.III, the information flow model and a measure
of system performance are introduced. The static and dynamic
routing strategies are described in Sec.IV and their performance
is analyzed in Sec.V.

\section{Network models}

\begin{figure}[ht]
 \subfigure{
  \includegraphics[scale=0.2]{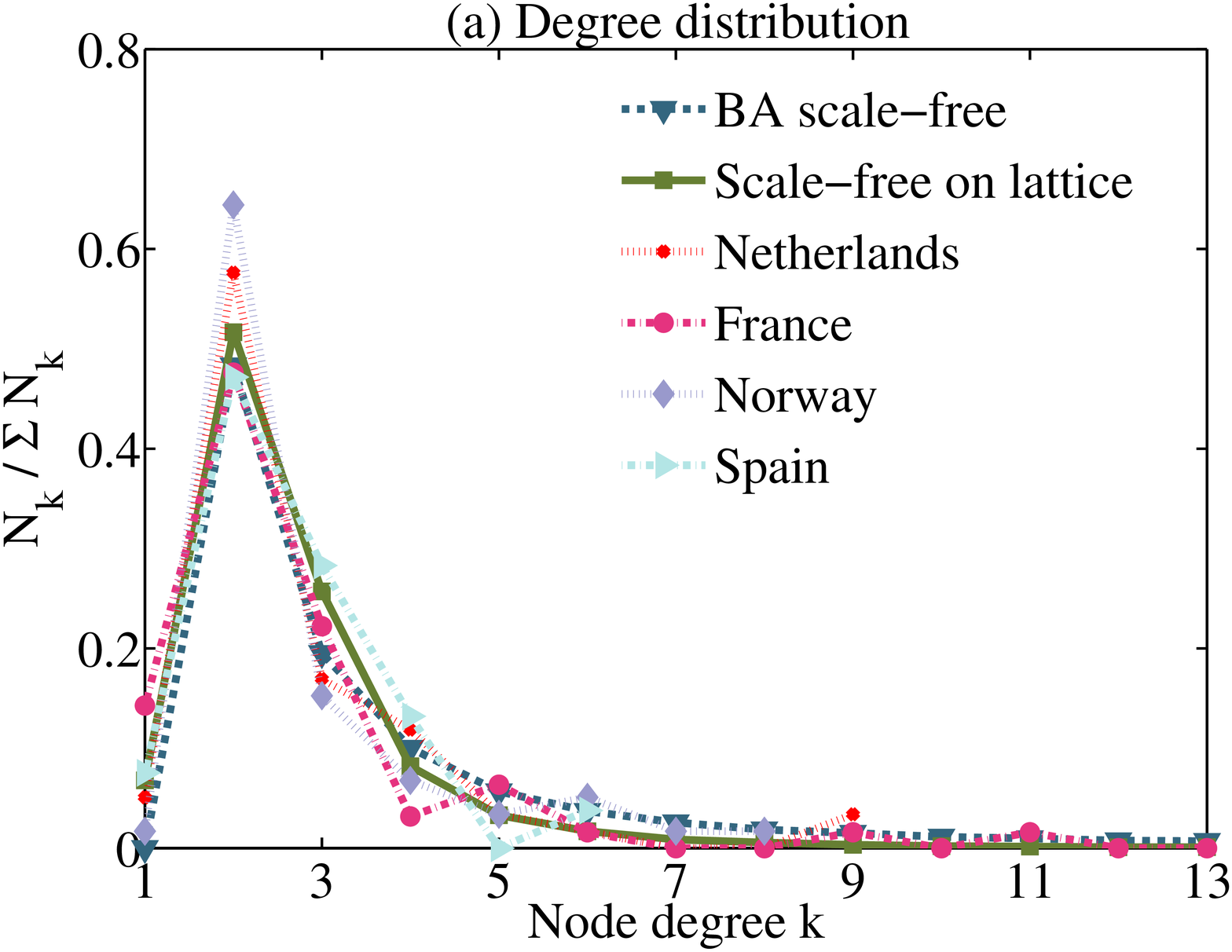} \label{degree}}
\subfigure{
  \includegraphics[scale=0.2]{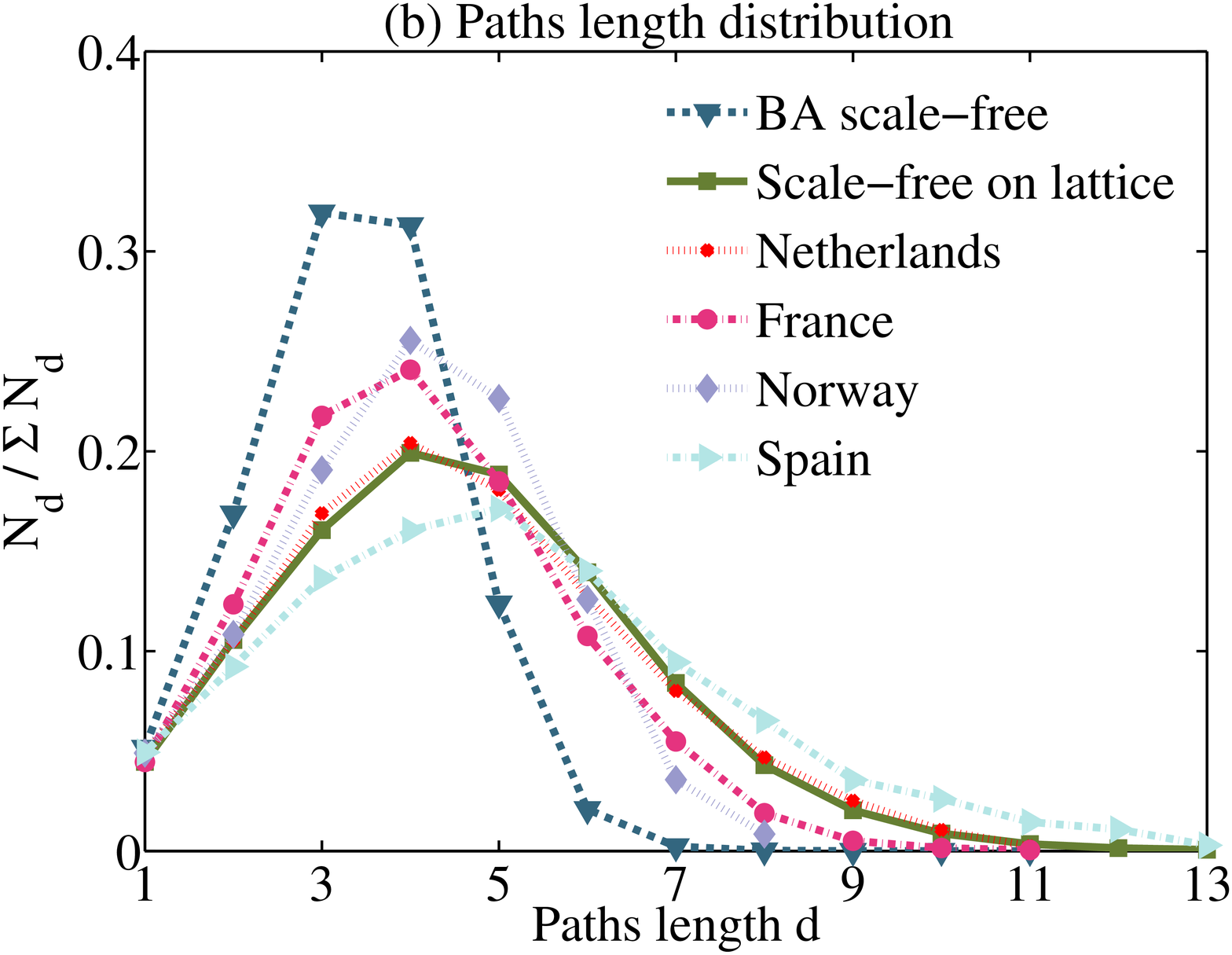} \label{path}}
\caption{Topological characteristics: (a) degree distribution; (b)
length of path probability density function for NRENs of the
Netherlands ($N=59$), France ($N=63$), Norway ($N=58$) and Spain
($N=53$) and generic network models: BA scale-free network
($N=64$, $m=2$) and scale-free network on lattice ($N=64$,
$\lambda=2$). }
\end{figure}

In this work, we compare topological network characteristics of
Barab\`asi-Albert scale-free model~\cite{BA}, the scale-free model
on lattice~\cite{rozenfeld} and the national research and
educational networks (NRENs) of the Netherlands, France, Norway
and Spain. Barab\`asi and Albert observed an existence of a high
degree of self-organization characterizing the large-scale
properties of complex networks~\cite{BA}. They have introduced a
model of the scale-free networks with two key elements:
probability that a new node connects to the existing nodes is not
uniform and there is a higher probability that it will be linked
to a node that already has a large number of connections. Thus,
Barab\`asi-Albert (BA) scale-free network model is formed in a
series of steps in which new nodes are incorporate into the
network. Algorithm is starting with a small number ($N_0$) of
nodes, and at every time step new nodes with $m$ connections are
added. To incorporate preferential attachment, the model assumes
that the probability of the new connection with the node $i$
depends on its connectivity $k_i$ and equals
$P(k_i)=k_i/\sum{k_j}$. After a few algorithm steps, distribution
of number of links per node takes scale-free form $P(k)\sim
k^{-\lambda}$. In this work, for $N_0=3$ and $m=2$, $n=61$
algorithm steps are preformed. Obtained network consists of $N=64$
nodes with $\lambda=2$. The obtained network degree distribution
corresponds well to the NRENs, cf. Fig.~\ref{degree}. The number
of nodes in different NRENs is in the case of the Netherlands
$N=59$, France $N=63$, Norway $N=58$ and Spain $N=53$.

However, the real-life networks are embedded into the geographical
space and constrained by the cost of the links between the nodes.
In the scale-free model on lattice, cf.~\cite{rozenfeld}, the
algorithm starts with a set of nodes that are identified with the
set of lattice vertices in an $M \times M$ square. The lattice
distance between two nodes is defined as minimal number of
"lattice steps" separating them in the regular lattice. In this
model, network nodes are randomly assigned with the number of
links ($k$) according to scale-free distribution
$P(k)=Ak^{-\lambda}$, $m \leq k < K$ and connected to its closest
neighbors. Therefore, exponent $\lambda$ is a model parameter. We
set $\lambda=2$, as obtained from the connectivity distribution of
NRENs, cf. Fig.~\ref{degree}. The choice of model parameter
$\lambda$ is also in accordance with the distribution of the
number of links per node obtained with BA model. Normalization
constant is $A \approx (\lambda-1)m^{\lambda-1}$.

At the first glance the most efficient mean to transfer
information through the network is along the shortest paths. The
distribution of the shortest path lengths is given in
Fig.~\ref{path}. The network diameter $D$, can be defined as the
maximal length of the shortest path between any two nodes in the
network, i.e., $D=\max \{d_{ij}\}$, where length of shortest path
from the node $i$ to the node $j$ is $d_{ij}$. The small network
diameter means that packets transmitted through the network,
travel from one node to another quickly along the shortest path.
As result the possibility of loss due to the congestion of the
transmitting nodes is reduced. From Fig.~\ref{path}, one can
observe that the path length distribution of the BA scale-free
network does not match NRENs well. This is not surprising, since
in the BA scale-free model the Euclidean distance between nodes is
irrelevant. For the version of scale-free network on lattice, the
model has desirable properties in terms of path lengths. Network
diameter in different NRENs is in the case of the Netherlands
$D=11$, France $D=11$, Norway $D=8$ and Spain $D=13$. Scale-free
network (BA) generation algorithm generates networks with
considerably smaller network diameters $D=6.84$. Average network
diameter obtained for scale-free on lattice model is $D=10.81$,
and compares well with diameters of NRENs.

\begin{figure}[ht]
\begin{center}
  \includegraphics[scale=0.2]{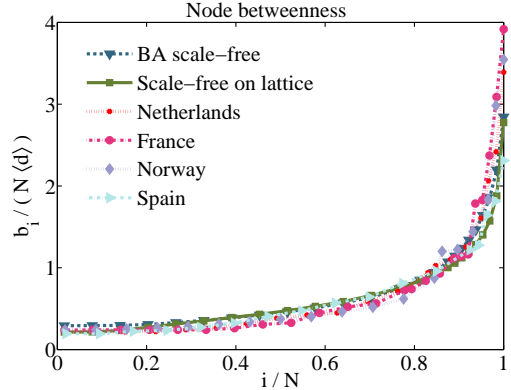}
\caption{Distribution of normalized node betweenness for NRENs of
the Netherlands ($N=59, \langle d \rangle = 4.54$), France ($N=63,
\langle d \rangle = 4.08$), Norway ($N=58, \langle d \rangle =
4.06$) and Spain($N=53, \langle d \rangle = 5.08$) and generic
network models: BA scale-free network ($N=64$, $m=2, \langle d
\rangle = 3.36$) and scale-free network on lattice ($N=64$,
$\lambda=2, \langle d \rangle = 4.59$). } \label{btw}
\end{center}
\end{figure}

The quality of the communication of the two non-adjacent nodes, i.e., node $j$
and node $k$, depends on the nodes belonging to the paths
connecting the nodes $j$ and $k$. Consequently, a measure of the
relevance of a given node for overall network performance can be
obtained by counting the number of geodesics going through it, and
defining the so-called node betweenness. More precisely, the
betweenness $b_i$ of the node $i$, sometimes referred to also as
load, is defined as~\cite{goh,havlin}:
\begin{equation}
 b_i=\sum_{j,k \in \mathcal{N},j \neq k}{n_{jk}(i)}
\end{equation}
where $n_{jk}(i)$ is the number of the shortest paths connecting
$j$ and $k$ and passing through the node $i$. Let $\langle d
\rangle$ denote the average path length of the given network
measured according to the shortest path routing rule. The
normalized betweenness distribution for NRENs of the Netherlands,
France, Norway and Spain and BA scale-free network and scale-free
network on latticeis is shown in Fig.~\ref{btw}. We observe that
both models reproduce betweenness characteristics of real networks
well.

\begin{figure}[ht]
 \subfigure{
  \includegraphics[scale=0.2]{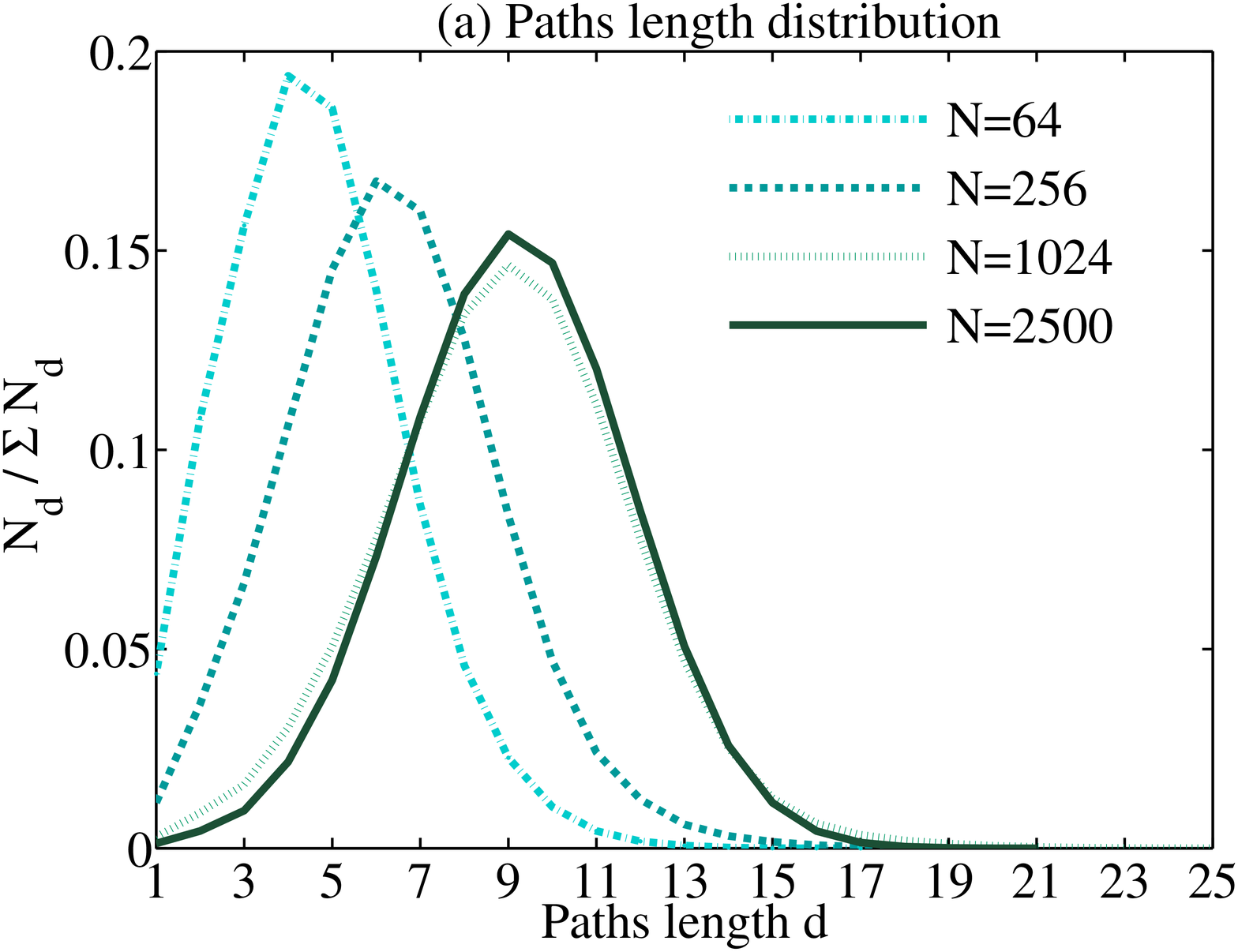} \label{SFL_SP}}
\subfigure{
  \includegraphics[scale=0.2]{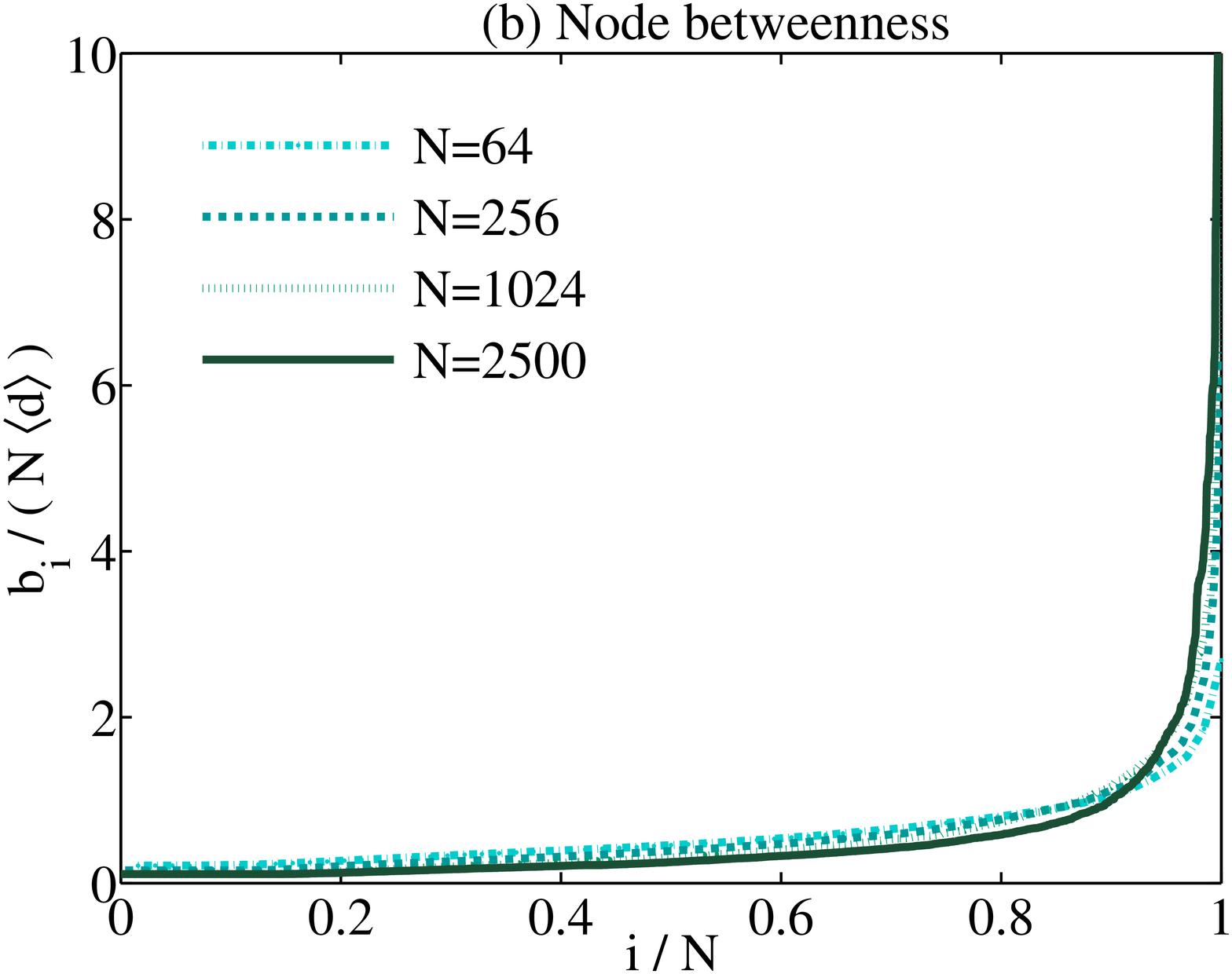} \label{SFL_btw}}
\caption{(a) Length of path probability density function for
scale-free network on lattice for various system sizes $N$. (b)
Distribution of normalized node betweeness for scale-free network
on lattice for various system sizes N, where $\langle d \rangle$
is average path length ($N=64,256,1024,2500$ and $\langle d
\rangle =4.59,6.38,8.98,9.14$ respectively). } \label{sys_size}
\end{figure}

In Fig.~\ref{sys_size} we examine the system-size dependent
($N=64,256,1024$ and $2500$) behavior of the path length and node
betweenness distribution for the scale-free on lattice model. The
average path length initially increases fast with the system size
(i.e., for $N<1024$). In the larger systems the path length
distribution changes little with the system size, cf. $N=1024$ and
$2500$ in Fig.~\ref{SFL_SP}. The reason for this is existence of
one or several nodes with high degree i.e. close to $N$, cf.
Ref.~\cite{cohen}. The average betweenness is roughly proportional
to the system size $N$ and the average path length $\langle
d\rangle$ as $\langle b_i \rangle\sim N\langle d \rangle$. Also,
the maximal value of node betweenness increases roughly with the
square of system size $N$, cf. Ref.~\cite{goh}.

\section{Information flow model}

In the information flow model all nodes are treated as both hosts
and routers. Each node has a predefined maximum packet routing
capacity $C$ and communication channels have an infinite capacity
to transmit the packets. If packets arrive to the node whose
routing capacity has been already reached (i.e., congested node),
they will be dropped. The dynamics of the model is as follows. At
each time step $t$, an information packet is created at random
node with the probability $p$. Therefore $p$ is the control
parameter: small values of $p$ correspond to the uncongested
(free) flow of packets and high values of $p$ correspond to the
high flow rate of packets. When a new packet is created, a
destination node, different from the origin, is chosen randomly in
the network. In this paper, we analyze the case that each node is
able to send one packet at each time step. The travel time $T$ of
a packet is defined as the time spent by the packet between its
source and destination. Here we do not take into account the time
delay of the information transfer at each node or link, so that
all data are delivered in a unit time, regardless of the distance
between any two nodes. Thus, during the following time steps $t+1,
t+2, \dots, t+T$, the packet travels toward its destination and
the time $T$ is related to the path length. Once the packet
reaches the destination node, it is delivered and disappears from
the network.

When the amount of packets is small, the network is able to
deliver all the packets that are generated. Conversely, when $p$
is large enough the number of generated packets is larger than the
number of packets that the network can manage to solve and the
nodes enter in a state of congestion. The characteristic that
measures the system performance is the packet drop probability
$\eta$,
\begin{equation}
 \eta=\frac{R_d}{R}
\end{equation}
defined as a ratio of the total number of deleted packets $R_d$
and the total number of generated packets $R$. A high drop
probability indicates that a large percentage of packets cannot
reach their destinations. Then, the quality of service is poorer.

\section{Routing strategies}

\subsection{Static routing}

Packets can be delivered according to different routing
strategies. When static weighted routing strategy is used, packets
choose the routing path with the minimum sum weight of links. For
any pair of source and destination node, there may be several
paths with the same weight between them. We randomly choose one of
these paths and put it into the fixed routing table which is
followed by all packets. It has be assumed that each node has the
same capability of delivering packets, that is, at each time step
all the nodes can deliver at most $C$ packets one step toward
their destinations according to the fixed routing table. Here we
compare two static routing strategies: $(i)$ the shortest path
routing, i.e., the links in the network have the same weight,
$w_{st}=1$, where $w_{st}$ is weight of link going from $s$ to
$t$. Routing communication along the shortest paths is of course
beneficial for speed, but if there is a limit to the node load and
network traffic is heavy, congestion is a threat to the nodes with
the largest betweenness. Obviously, bypassing high-degree nodes,
packet will have more chance to reach its destination. $(ii)$ In
the second, efficient path routing strategy, the weight of the
link between the nodes $s$ and $t$ is defined as~\cite{yan}:
\begin{equation}
w_{st}=\left(\frac{k_{s}+k_{t}}{\displaystyle\min_{i \neq
j}{(k_{i}+k_{j})}}\right)^\beta \label{lcost},
\end{equation}
where $k_i$ denotes degree of node $i$ and $\beta$ is an
adjustable parameter. The efficient path between nodes $i$ and $j$
is corresponding to the route that makes the sum weight of links
minimum. As for any pair of source and destination, there may be
several efficient paths between them. We randomly choose one of
them and put it into the fixed routing table which is followed by
all the information packets.

\begin{figure}[ht]
 \subfigure{
  \includegraphics[scale=0.2]{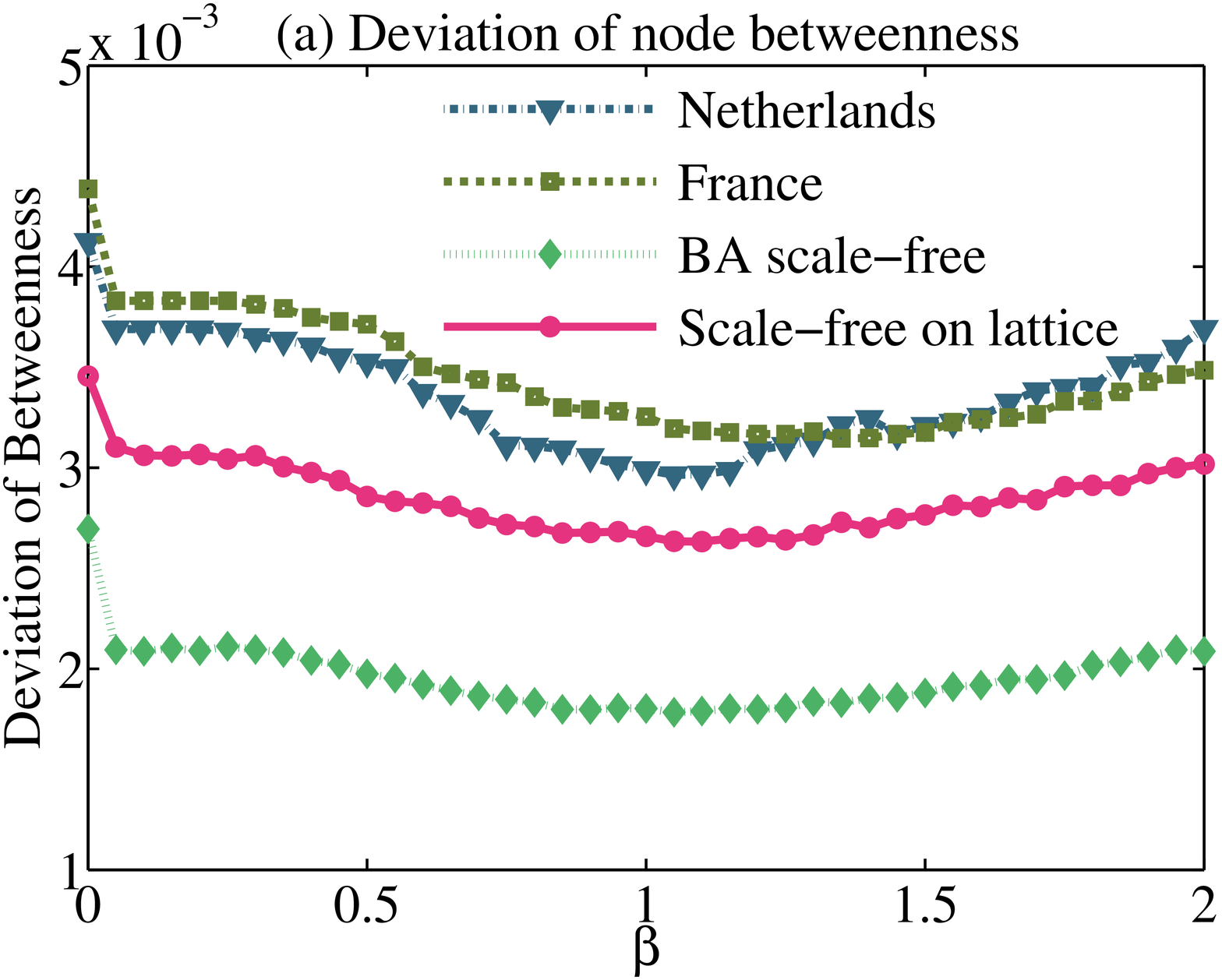} \label{btw_beta}}
\subfigure{
  \includegraphics[scale=0.2]{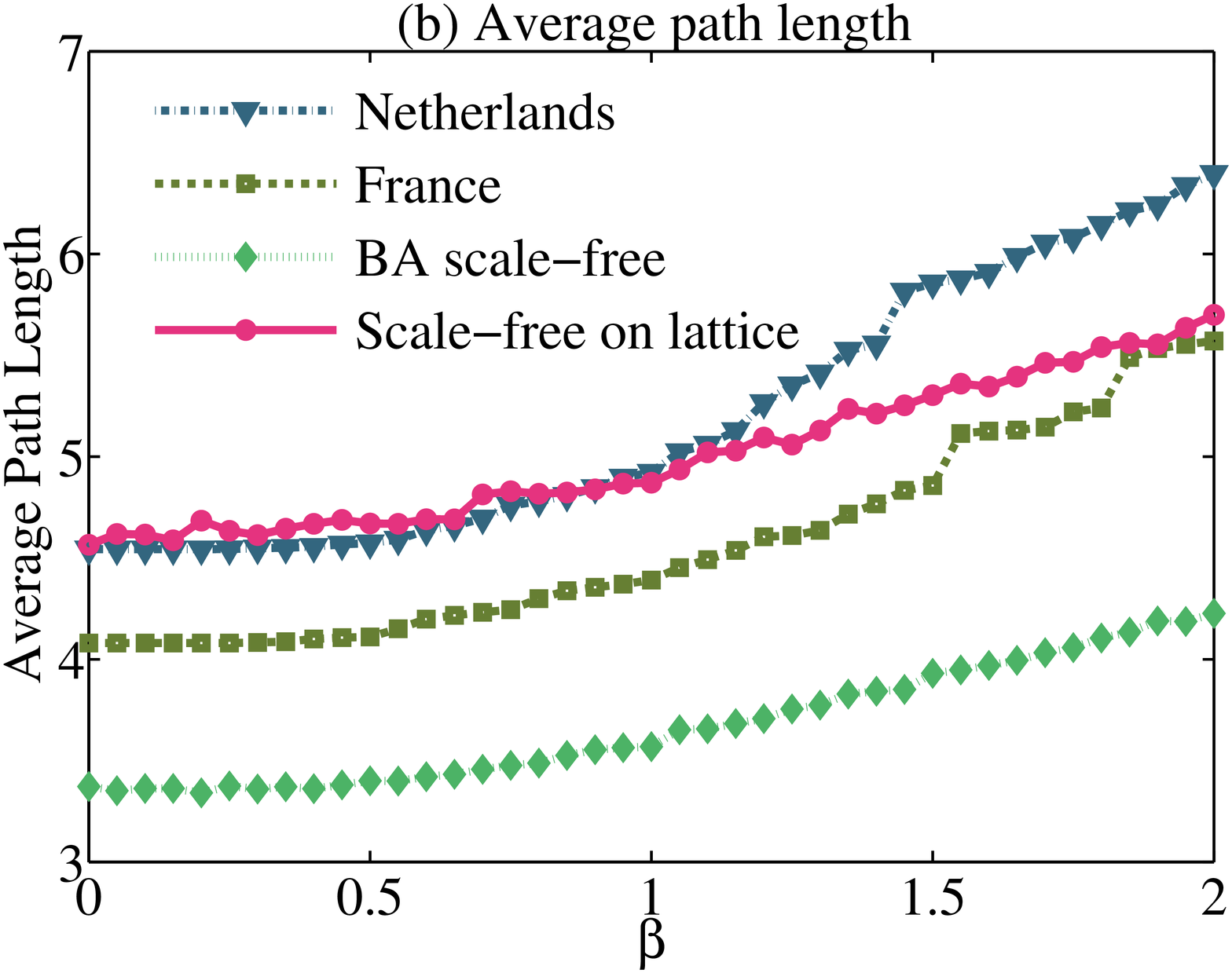} \label{path_beta}}
\caption{(a) Deviation of node betweenness vs parameter $\beta$ from Eq.~\ref{lcost}
and (b) average path length vs parameter $\beta$ for different network models.}
\end{figure}

In Fig.~\ref{btw_beta}, we show the betweenness deviation as a
function of $\beta$ on different scale-free networks. The optimal
routing strategy for generic scale-free network and scale-free
network on lattice is corresponding to $\beta_{opt}=1 \pm 0.1$,
where betweenness deviation is the smallest. This is also optima
value for NREN of the Netherlands. The optimal value for French
NREN is higher, i.e., $\beta_{opt}=1.2\pm 0.1$, due to larger
betweenness deviation. Obviously $\beta = 0$ recovers the shortest
path length. In comparison with the shortest path routing strategy
average path length slightly increases with $\beta$, cf.
Fig.~\ref{path_beta}.

\subsection{Dynamic routing}

In this paper we also analyze the effectiveness of the dynamic
deflection routing strategy. We assume each node has a knowledge
about the load of its neighbors. If a package is about to arrive
to a congested node, it will be deflected, i.e., its path will be
dynamically extended. In other words, if we denote a path between
nodes $i$ and $j$ as $P(i \rightarrow j):=x_0,x_1, \dots,
x_{n-1},x_n$, where $x_0=i$ and $x_n=j$. The node $x_m$, instead
forwarding packet to the congested node $x_{m+1}$, deflects the
packet back to the node $x_{m-1}$. In the moment $t+1$, the node
$x_{m-1}$ sends the packet to the node $x_m$, and in the moment
$t+2$ the packet is sent to the node $x_{m+1}$. The deflection of
the packet is tried only once.

The total number of the packets in network is equal to $p N
\langle d \rangle$. Since node betweenness represents the number
of paths going through the node, the probability that a packet in
one time step is going through node $i$ is $B_i =
\frac{b_i}{N(N-1)\langle d \rangle}$. Therefore, average number of
packets going through node $i$ per step is $q_i=p N \langle d
\rangle \times B_i = p \frac{b_i}{N-1}$. In order to avoid a
reduction of the system performance due to the deflection traffic,
we introduce a condition that a node can send packet back only if
$q_i<0.5C$. The localized congestion events are more likely to
occur at the nodes with higher betweenness and that in the regime
when traffic is not heavily congested. We find that it is
sufficient to implement the deflection routing only in the nodes
with highest betweenness (about $10\%$ of all nodes). A higher
amount of the nodes with deflection routing capability does not
improve significantly the network performance.

\section{Simulation results}

\begin{figure}[ht]
\begin{center}
 \includegraphics[scale=0.3]{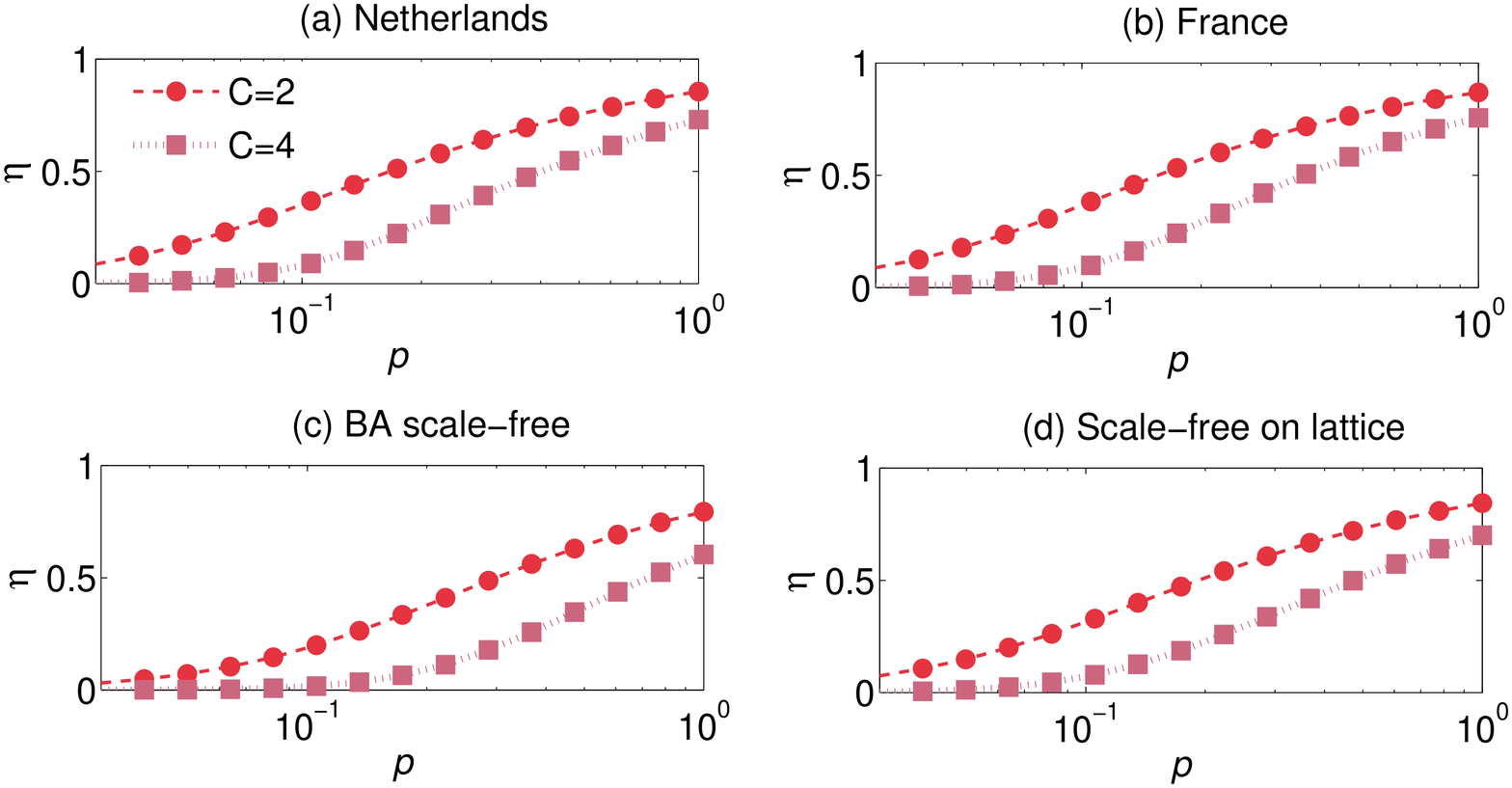}
\caption{ Packet drop probability $\eta$ vs. the
packet-generating rate $p$ for the shortest path routing
algorithm. We have set node transfer capacity $C=2$ (circles)
and $C=4$ (squares). \label{eta}}
\end{center}
\end{figure}

To compare different routing strategies, we apply previously
described routing algorithms and measure packet drop probability
$\eta$ in the two generic networks and NRENs of France and the
Netherlands. As shown in Fig.~\ref{eta}, the scale-free network on
lattice under the shortest path routing strategy reproduces well
behavior of both NREN topologies, for both simulated node transfer
capacities $C=2$ and $4$. We can observe that in the real-world
networks and scale-free network on lattice for $p>0.1$ and for
node capacity $C=2$ more than $40\%$ of generated packets is
dropped. Also, when amount of the traffic in the network is small
($p<0.1$), there is a significant amount of loss, i.e., more than
$5\%$. For the higher node transfer capacity ($C=4$) number of
deleted packets significantly decreases and there are almost no
packets loss when the network traffic is small. The BA scale-free
network is less prone to the congestion compared to the scale-free
network on lattice and two NRENs, which is expected since this
network has a shorter average path length.

\begin{figure}[ht]
\begin{center}
 \includegraphics[scale=0.3]{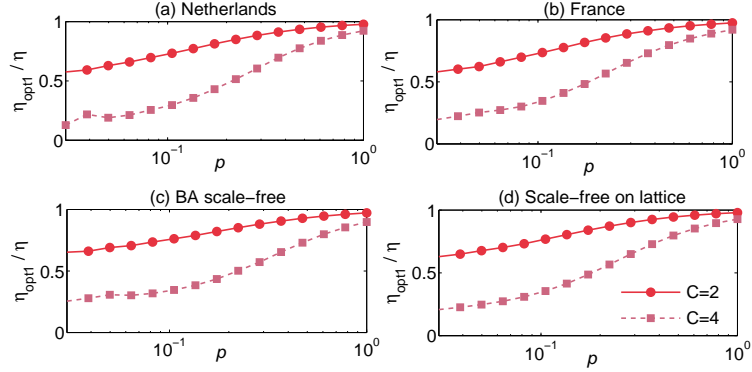}
\caption{Comparison between packet drop probability $\eta$ as a
function of packet-generating rate $p$ for the shortest path
routing and static weighted routing strategy with $\beta_{opt}$
($\eta_{opt1}$). We have set node transfer capacity $C=2$
(circles) and $C=4$ (squares). \label{eta1}}
\end{center}
\end{figure}

\begin{figure}[ht]
\begin{center}
 \includegraphics[scale=0.3]{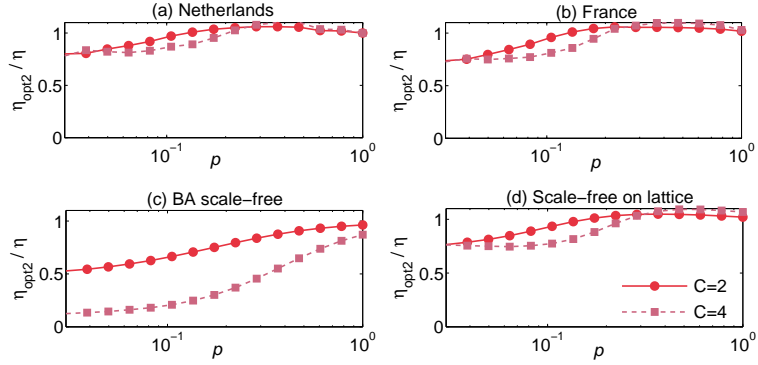}
\caption{Comparison between packet drop probability $\eta$ as a
function of packet-generating rate $p$ for the shortest path
routing and dynamic routing algorithm ($\eta_{opt2}$). We have set
node transfer capacity $C=2$ (circles) and $C=4$ (squares).
\label{eta2}}
\end{center}
\end{figure}

\begin{figure}[ht]
\begin{center}
 \includegraphics[scale=0.3]{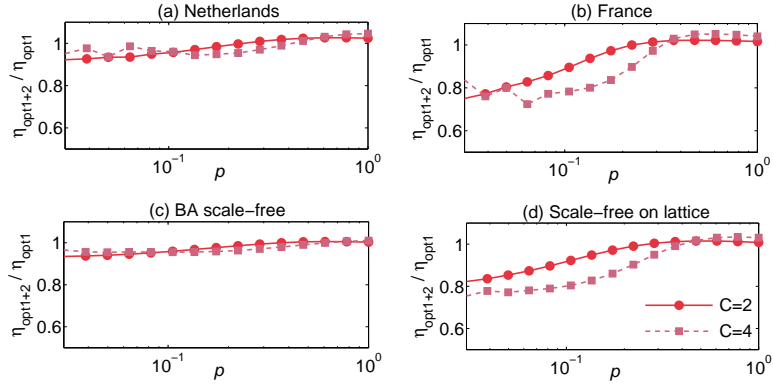}
\caption{Comparison between packet drop probability $\eta$ as a
function of packet-generating rate $p$ for static weighted routing
strategy, $\beta_{opt}$ ($\eta_{opt1}$) and combined static and
dynamic routing ($\eta_{opt1+2}$). We have set node transfer
capacity $C=2$ (circles) and $C=4$ (squares). \label{eta3}}
\end{center}
\end{figure}

\begin{figure}[ht]
\begin{center}
 \includegraphics[scale=0.2]{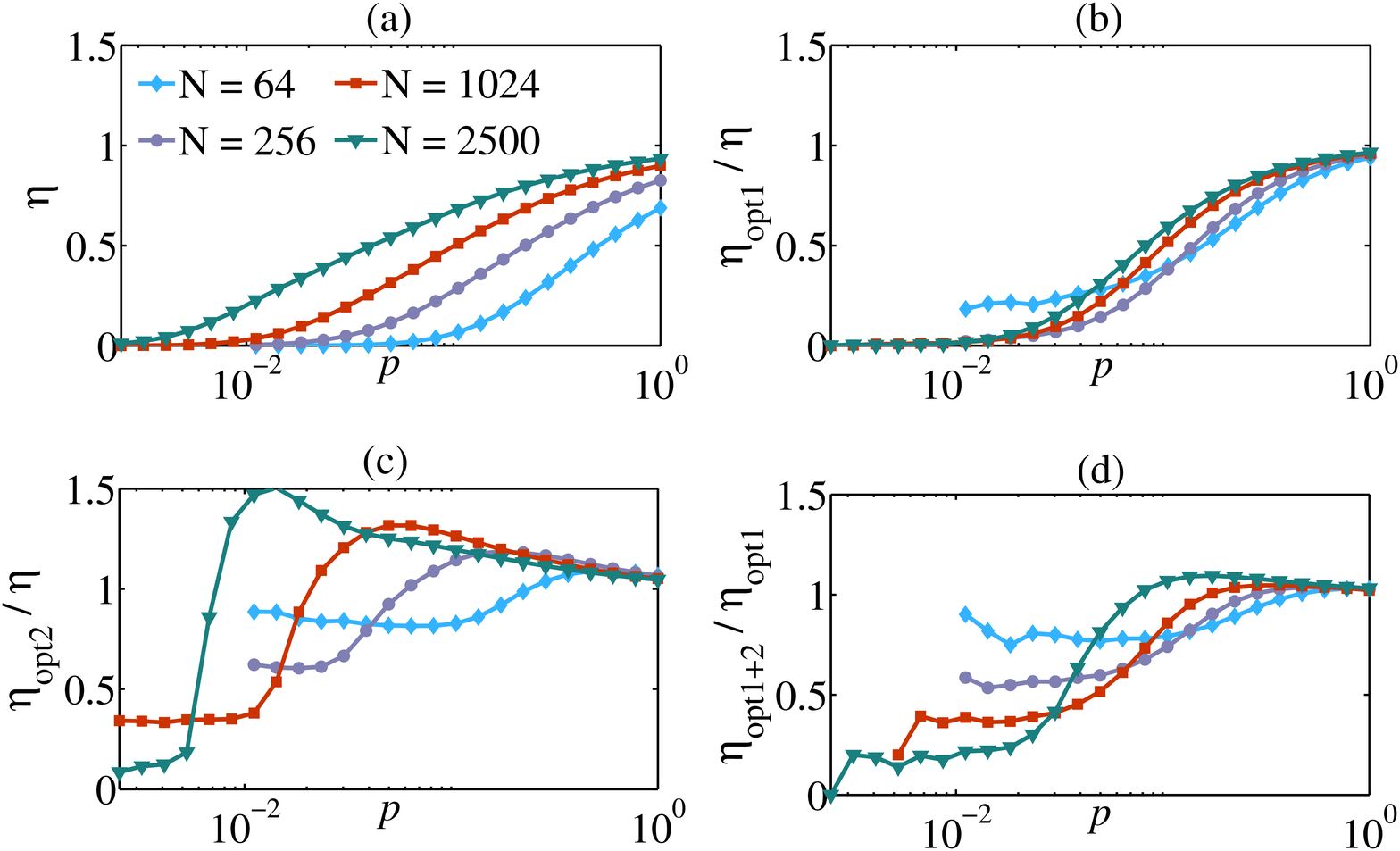}
\caption{ Packet drop probability ($\eta$) in four scale-free on
lattice networks with different sizes $N=64,256,1024$ and $2500$.
Results for different routing strategies are compared: (a) the
shortest path routing algorithm, (b) static weighted routing
strategy and shortest path routing ($\eta_{opt1}/\eta$), (c)
dynamic routing algorithm and shortest path routing
($\eta_{opt2}/\eta$), and (d) combined static and dynamic routing
strategy and static weighted routing strategy
($\eta_{opt1+2}/\eta_{opt1}$). The value of node transfer capacity
used in this figure is fixed ($C=4$). \label{eta4}}
\end{center}
\end{figure}

Impact of the static and dynamic routing strategies is analyzed in
Fig.~\ref{eta1} and Fig.~\ref{eta2}, respectively. The results
confirm that the static weighted routing using the topological
information can greatly improve the traffic flow for small $p$
values ($p < 0.1$), in comparison with the shortest path routing
mechanism. The static weighted routing method proves to be better
for NREN topologies and scale-free model on lattice. The dynamic
routing strategy is better in case of BA scale-free networks due
to shorter paths, cf. Fig.~\ref{eta2}. The shorter paths between
nodes result in a lower amount of the packages being transferred
through the network at any instance of time compared to other
studied networks, allowing for the dynamic deflection algorithm to
be more efficient.

The dynamic and static routing strategies are complementary and
can be combined. The system performance for combined static and
dynamic routing and the static routing is compared in
Fig.~\ref{eta3}. The redistribution of the traffic load from the
most congested nodes enables more efficient the dynamic deflection
routing. Improvement is in range from $5\%-20\%$. In NREN of
Netherlands four most congested nodes after path redistribution
where in the proximity, and combined effect of dynamic-static
routing strategy is rather weak, cf. Fig~\ref{eta3}(a). On the
other hand, the French NREN shows a considerable improvement due
to the larger distance between the most congested nodes, cf.
Fig~\ref{eta3}(b). This is consequence of the fact that both
studied networks are small and the difference in the performance
of the routing algorithms is a result of the variations in the
topology. The scale-free model on lattice allows us to evaluate
and compare routing strategy on large number of artificially
generated networks. In this way we obtain information, how whole
class of the networks responds to the routing strategy in average
and we observe a considerable improvement in performance, cf.
Fig.~\ref{eta3}(d).

Until now, we have tested the routing strategies for small
networks (with $N\leq64$ nodes). At this point, we want to show
how the static and dynamic strategy behave in the larger networks,
i.e., $N=256, 1024,$ and $2500$. In Fig.~\ref{eta4}, we show the
relationship between packet drop probability $\eta$ and network
size $N$ under different routing strategies on scale-free networks
on lattice. In the case of the shortest path routing, the results
are consistent with our intuition: with the increase of the system
size $N$ the congestion starts at smaller packet generation rates
$p$. In order to understand where the packet drop first occurs and
now congestion develops, we need to look into the properties of
the single network nodes. The probability of information drop
(loss) on certain node depends on number of the packets going
through it. The average number of packets going through node $i$
per step is $q_i= p \frac{b_i}{N-1}$ and the largest values of
node betweenness, $b_{max}$, increases faster then linear with
system size, cf. Ref.~\cite{goh} and Fig.~\ref{SFL_btw}. Therefore
increase of system size $N$ results in increase of drop
probability. Fig.~\ref{eta4}(b) compares the relation of the
packet drop probability $\eta$ vs packet-generating rate $p$ under
the static weighted routing and the shortest path routing with
different network size $N$. The efficiency of static weighted
routing strategy increases with the system size $N$. Also, the
larger networks compared to very small ones, i.e., $N=64$, offer
more possibility to redistribute load and therefore the strategy
is more efficient. On the other hand, dynamic routing algorithm
changes alone is less efficient in larger networks, see
Fig.~\ref{eta4}(c). The reason for this is existence of the nodes
with the large node degree (comparable to size of the network),
cf. Ref.~\cite{cohen}. As result the number of deflected packets
increases with the network size increasing congestion in the
surrounding nodes. This problem is alleviated in the combination
of the static weighted and dynamic routing algorithms, cf.
Fig.~\ref{eta4}(d). Since in this case the most congested nodes
are avoided and the load is more evenly distributed, the dynamic
routing can therefore further reduce congestion. We observe that
improvement of network performance due to combined dynamic-static
routing strategy is increasing with system size.

\section{Conclusions}

In summary, we have introduced a information flow model for
networks without buffering capacity. We have shown that scale-free
model on lattice reproduces well both topological and information
transport characteristics of the small national research and
educational networks of the Netherlands, France, Norway and Spain.
In the small networks, the point of the network congestion and
amount of information lost strongly depends on the underlying
network structure. We have further described a dynamic deflection
routing strategy suitable for the networks without buffers. The
proposed strategy dynamically extends packet path before it
reaches a congested node. Simulations on two real world networks
and two generic networks with small diameter confirmed the highest
traffic capacity under combined static and dynamic routing is
achieved when compared with the shortest path routing strategy and
the static weighted routing strategy. Moreover, we found that the
dynamic deflection routing can further improve network information
transport capacity when combined with the efficient path routing
strategy both in small and large networks.

\section*{Acknowledgements}

Discussions with Jovan Radunovi\'{c} and Vladica Tintor helped to motivate this effort.
The authors acknowledge support by the Ministry of Science of the
Republic of Serbia, under project No. ON171017. Numerical
simulations were run on the AEGIS e-Infrastructure,
supported in part by FP7 projects EGIInSPIRE, PRACE-1IP,
PRACE-2IP, and HP-SEE. The authors also acknowledge support
received through SCOPES grant IZ73Z0-128169 of the Swiss National
Science Foundation.

%%%%%%%%%%%%%%%% REFERENCES

\newpage

%\clearpage


\begin{thebibliography}{widestlabel}

\bibitem{kim1} D.-H. Kim and A. E. Motter, J. Phys. A: Math. Theor. 41, 224019 (2008).
\bibitem{kim2} D.-H. Kim and A. E. Motter, New J. Phys. 10, 053022 (2008).
\bibitem{danila1} B. Danila, Y. Yu, J. A. Marsh, and K. E. Bassler, Phys. Rev. E 74, 046106 (2006).
\bibitem{danila2} B. Danila, Y. Sun, and K. E. Bassler, Phys. Rev. E 80, 066116 (2009).
\bibitem{yan} G. Yan, T. Zhou, B. Hu, Z.-Q. Fu, and B.-H. Wang, Phys. Rev. E 73, 046108 (2006).
\bibitem{tang} M. Tang and T. Zhou, Phys. Rev. E 84, 026116 (2011).
\bibitem{guimera} R. Guimer\`a, A. Arenas, A. D\'iaz-Guilera, F. Giralt, Phys. Rev. E 66, 026704 (2002).
\bibitem{wang1} W.-X. Wang, B.-H. Wang, C.-Y. Yin, Y.-B. Xie, and T. Zhou, Phys. Rev. E 73, 026111 (2006).
\bibitem{yin} C.-Y. Yin, B.-H. Wang, W.-X. Wang, T. Zhou, and H.-J. Yang, Phys. Lett. A 351, 220 (2006).
\bibitem{sreenivasan} S. Sreenivasan, R. Cohen, E. Lopez, Z. Toroczkai, and H. E. Stanley, Phys. Rev. E 75, 036105 (2007).
\bibitem{pu} C.-L. Pu, S.-Y. Zhou, K. Wang, Y.-F. Zhang, and W.-J. Pei, Physica A 391, 866 (2012).
\bibitem{wang2} W.-X. Wang, C.-Y. Yin, G. Yan, and B.-H. Wang, Phys. Rev. E 74, 016101 (2006).
\bibitem{kujawski} B. Kujawski, G. J. Rodgers, and B. Tadi\'{c}, Lect. Notes Comput. Sci. 3993, 1024 (2006).
\bibitem{echenique1} P. Echenique, J. G\'omez-Garde\H{n}es, and Y. Moreno, Phys. Rev. E 70, 056105 (2004).
\bibitem{echenique2} P. Echenique, J. G\'omez-Garde\H{n}es, and Y. Moreno, Europhys. Lett. 71, 325 (2005).
\bibitem{ling} X. Ling, M.-B. Hu, R. Jiang, and Q.-S. Wu, Phys. Rev. E 81, 016113 (2010).
\bibitem{tintor} V. Tintor and J. Radunovi\'{c}, Photonic Network Communications 18, 55 (2009).
\bibitem{wischik1} D. Wischik, In ECOC, Scotland (2005).
\bibitem{wischik2} D. Wischik and N. McKeown, ACM CCR 35(2), 75 (2005).
\bibitem{raina} G. Raina, D. Towsley, and D. Wischik, ACM CCR 35(2), 79 (2005).
\bibitem{enachescu1} M. Enachescu et al., ACM CCR 35(2), 83 (2005).
\bibitem{enachescu2} M. Enachescu et al., In IEEE INFOCOM, Spain (2006).
\bibitem{gorinsky1} S. Gorinsky, A. Kantawala, and J. Turner, In ISCC, Spain (2005).
\bibitem{gorinsky2} S. Gorinsky, A. Kantawala, and J. Turner, Simulation 83(3), 245 (2007).
\bibitem{wong} E. W. M. Wong, L. L. H. Andrew, T. Cui, B. Moran, A. Zalesky, R. S. Tucker, M. Zukerman,
J. Lightwave Technol. 27(14), 2817 (2009).
\bibitem{BA} A.-L. Barab\`asi, R. Albert, Science 286, 509 (1999).
\bibitem{dorogovtsev} S.N. Dorogovtsev, J.F.F. Mendes, {\it Evolution of Networks}, Oxford University Press, Oxford, 2003.
\bibitem{rozenfeld} A.F. Rozenfeld, R. Cohen, D. ben-Avraham, S. Havlin, Phys. Rev. Lett. 89, 218701 (2002).
\bibitem{surfnet} http://www.surfnet.nl.
\bibitem{renater} http://www.renater.fr.
\bibitem{uninett} https://www.uninett.no.
\bibitem{rediris} http://www.rediris.es.
\bibitem{dijkstra} E. W. Dijkstra, Numer. Math. 1, 269 (1959).
\bibitem{goh} K.-I. Goh, B. Kahng, and D. Kim, Phys. Rev. Lett. 90, 058701 (2003).
\bibitem{havlin} S. Havlin and R. Cohen, {\it Complex Networks - Structure, Robustness and Function}, Cambridge University Press, Cambridge, 2010.
\bibitem{cohen} R. Cohen and S. Havlin, Phys. Rev. Lett. 87, 278701 (2001).
\end{thebibliography}
\end{document}